\pdfoutput=1

\documentclass[final,5p,times,twocolumn,sort&compress]{elsarticle}

\usepackage{amsmath}
\usepackage{amsfonts}
\usepackage{amssymb}

\usepackage{graphicx} 
\usepackage{dcolumn}  
\usepackage{subfigure} 

\begin{document}
\begin{frontmatter}

\title{An algorithm for quantifying dependence in multivariate data sets}

\author{M.~Feindt}
\author{M.~Prim\corref{cor1}}\ead{michael.prim@kit.edu}
\cortext[cor1]{Corresponding author. Tel.: +49 721 608-43418; Fax: +49 721 608-47930}
\address{Institut f\"ur Experimentelle Kernphysik, Karlsruher Institut f\"ur Technologie, Campus S\"ud, Postfach 69 80, 76128 Karlsruhe}

\begin{abstract}
We describe an algorithm to quantify dependence in a multivariate data set. The algorithm is able to identify any linear and non-linear dependence in the data set by performing a hypothesis test for two variables being independent. As a result we obtain a reliable measure of dependence.

In high energy physics understanding dependencies is especially important in multidimensional maximum likelihood analyses. We therefore describe the problem of a multidimensional maximum likelihood analysis applied on a multivariate data set with variables that are dependent on each other. We review common procedures used in high energy physics and show that general dependence is not the same as linear correlation and discuss their limitations in practical application.

Finally we present the tool CAT, which is able to perform all reviewed methods in a fully automatic mode and creates an analysis report document with numeric results and visual review.
\end{abstract}


\end{frontmatter}

\section{Introduction}
\label{sec:introduction}
This paper describes an algorithm for quantifying dependencies in a multivariate data set. Throughout this paper we will, in contrast to common jargon, strictly speak of correlation only in the context of linear correlation, whereas dependence is used for general, linear and also non-linear, correlation. Understanding dependencies is especially useful and necessary in multidimensional likelihood analysis, a technique widely used in high energy physics (HEP). Such analysis entails constructing a probability density function (PDF) describing the multivariate data set. In many analyses dependencies among different variables are neglected in the PDF. It is required to somehow prove that neglecting the dependencies is a valid procedure as e.\,g. they are small.

In section \ref{sec:maxlikelihood} a brief introduction of the maximum likelihood method is given to illustrate the problems that arise from a data set with variables that are not independent. Sections \ref{sec:linearcorrel} and \ref{sec:projections} will review existing methods and discuss their limitations. In section \ref{sec:correlana} a new algorithm for quantifying dependence is explained and section \ref{sec:CAT} presents CAT, a fully automatic analysis tool. Section \ref{sec:howto} will briefly outline which possibilities exist to deal with dependencies in the data set.

\section{Maximum likelihood analysis}
\label{sec:maxlikelihood}
Consider an unbinned extended maximum likelihood analysis of a data set with events of different categories $c$ (e.\,g. signal and background). The log-likelihood function is expressed as:
\begin{align}
\ln \mathcal{L} = \sum_{j = 1}^{N} \ln \left\lbrace \sum_{i = 1}^{N_c} N_i \mathcal{P}_i(\vec{x}_j) \right\rbrace - \sum_{i = 1}^{N_c} N_i,
\end{align}
where
\begin{itemize}
\item $N$ is the total number of events in the data set,
\item $N_c$ is the number of different categories in the data set,
\item $N_i$ is the expected number of events for the $i$\textsuperscript{th} category,
\item $\mathcal{P}_i$ is the PDF for the $i$\textsuperscript{th} category,
\item $\vec{x}_j$ is the $n$-dimensional vector of variable values for the $j$\textsuperscript{th} event.
\end{itemize}

In the analysis the log-likelihood is maximized by changing the $N_i$ yields to extract the most likely set. If $\vec{x}$ has more than one dimension, one usually speaks of a multidimensional analysis.

The crucial point of a maximum likelihood analysis is to choose the model properly. Such model might be either provided by theory or must be derived from simulated data and sideband studies. The latter is a common practice in HEP. In case of a multidimensional analysis the model must also describe the dependencies among different variables correctly. If no theoretical model exists, e.\,g. for combinatorial background components, experimentalists usually start by describing the $n$-dimensional PDF as a product of marginal distributions: 
\begin{align}
\label{eqn:independence}
\mathcal{P}_i(\vec{x}) = \mathcal{M}_1(x_1) \times \mathcal{M}_2(x_2) \times \cdots \times \mathcal{M}_n(x_n).
\end{align}
This procedure is entirely valid with no dependencies between different variables. Indeed equation \ref{eqn:independence} is the definition of independence among variables.

If such a model $\mathcal{P}_i$ shall be used for the $i$\textsuperscript{th} category, the $N_i$ events must have no dependencies among the different variables. It is the task of the experimentalist to prove that this assumption is valid and it is the aim of the following sections to provide assistance.

\section{Linear correlation coefficient}
\label{sec:linearcorrel}
\begin{figure*}[t]
  \centering
  \subfigure[]{
    \includegraphics[width=0.3\linewidth, angle=0]{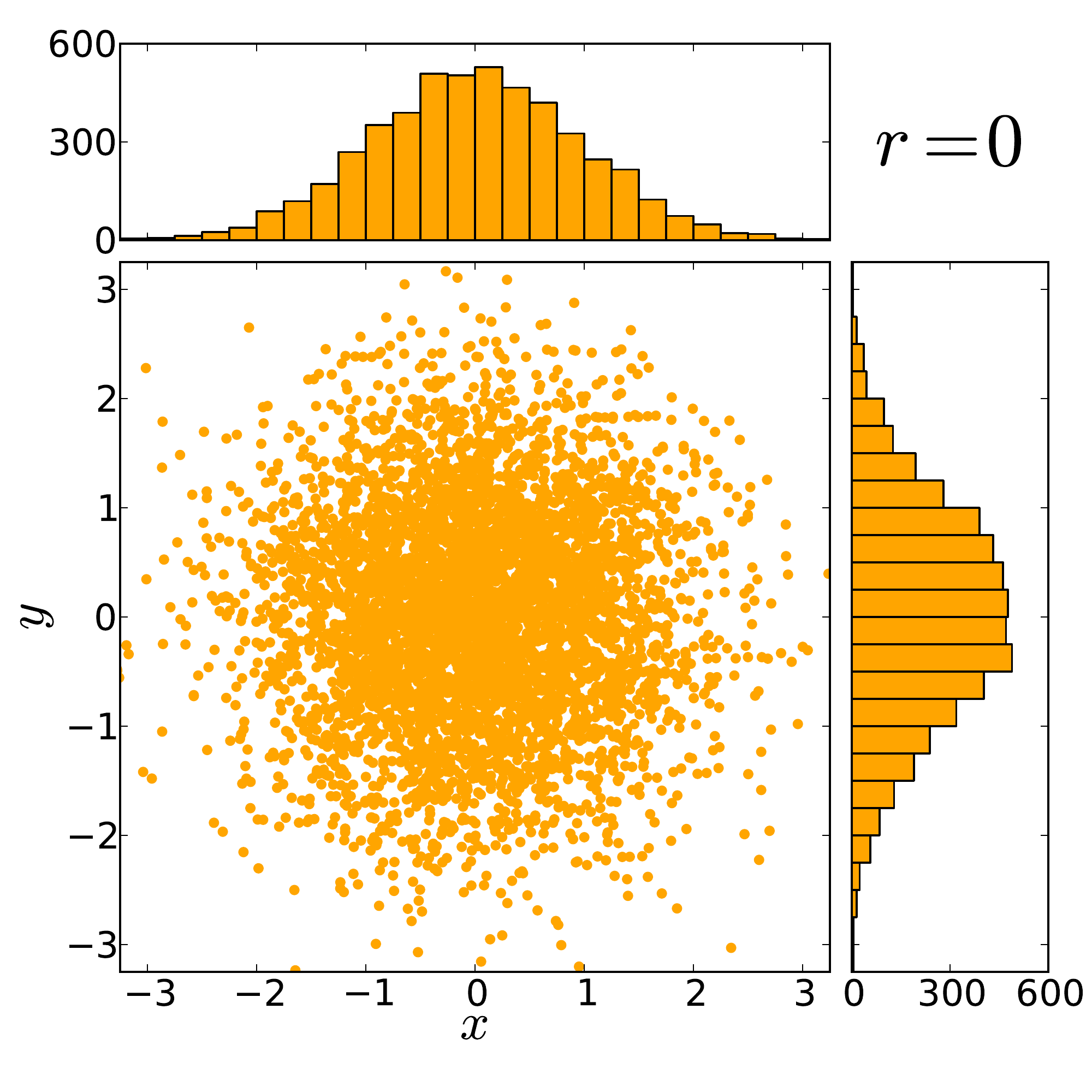}
    \label{fig:pearson0}
  }
  \subfigure[]{
    \includegraphics[width=0.3\linewidth, angle=0]{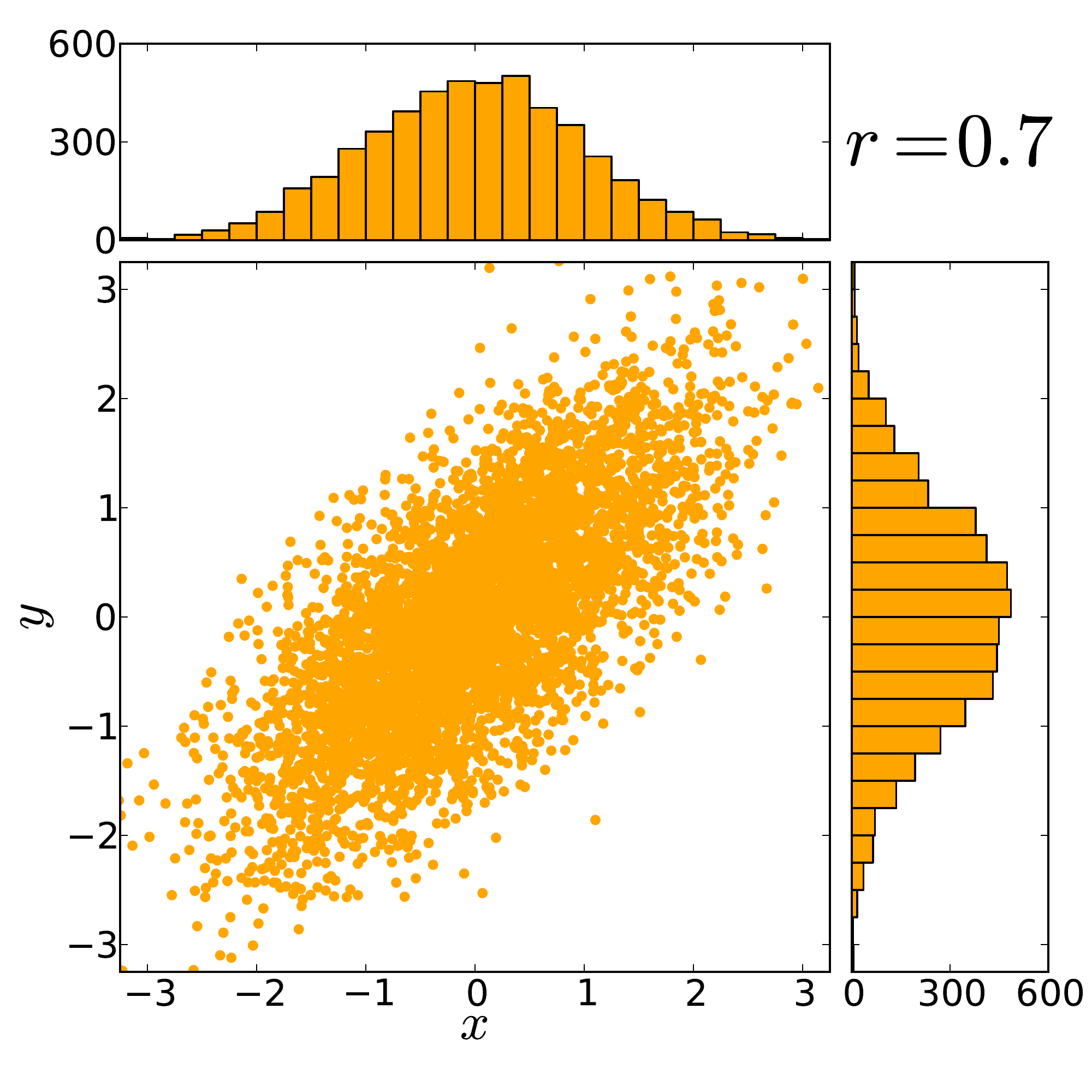}
    \label{fig:pearson1}
  }
  \subfigure[]{
    \includegraphics[width=0.3\linewidth, angle=0]{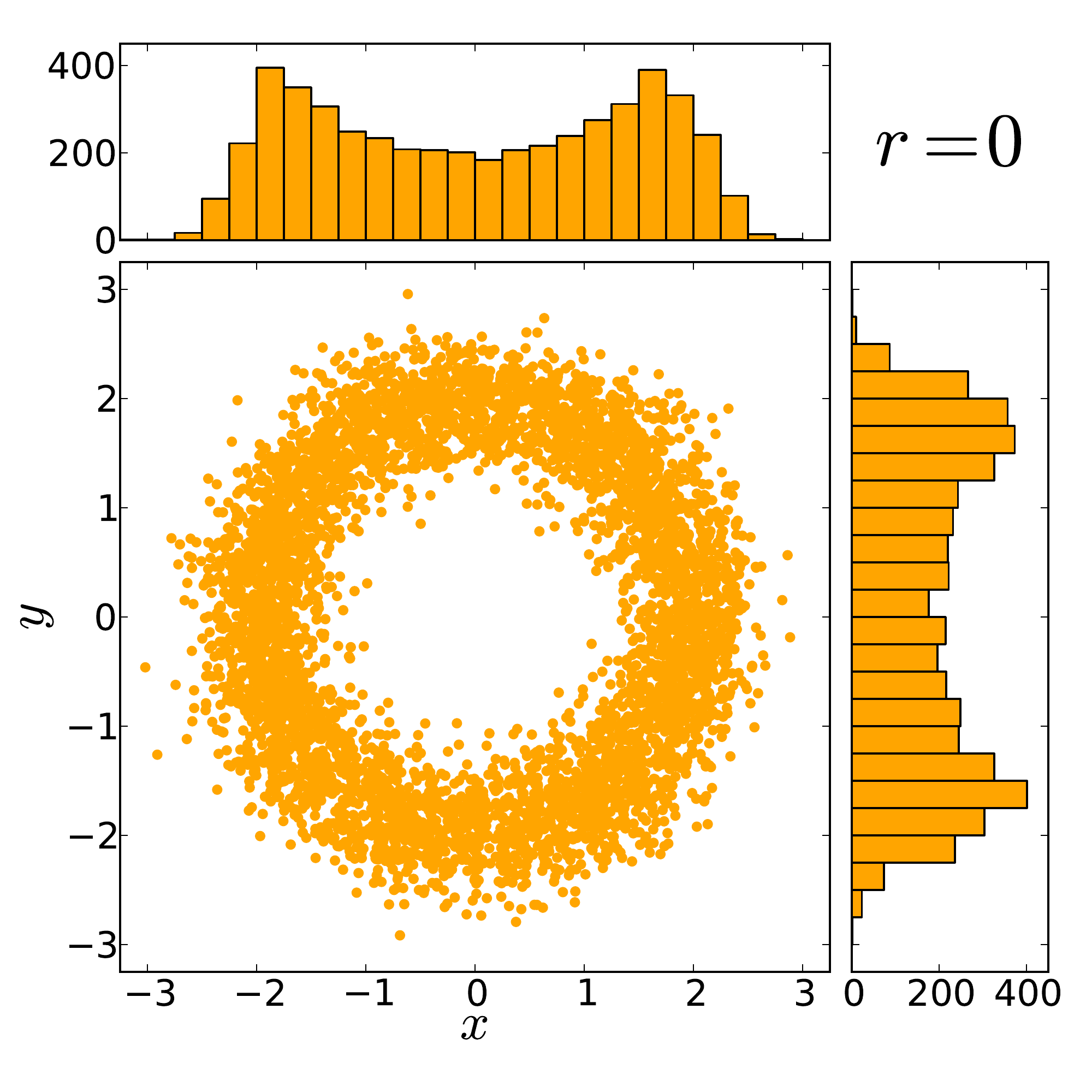}
    \label{fig:pearson2}
  }
  \caption{Example for two variables $x$ and $y$ following a Gaussian normal distribution with no (a) and 70\% linear correlation (b). Example of circular distributed variables $x$ and $y$  is shown in (c). Marginal distribution $P(x)$($P(y)$) of each sample is shown above (right) of the scatter plot.}
  \label{fig:pearson}
\end{figure*}

One often used quantity to describe dependence among two variables $x$ and $y$ is the linear correlation coefficient $r$. For a given sample of $N$ events, it can be computed from the data by
\begin{align}
\label{eqn:pearson}
r = \frac{\sum_{i=1}^{N} (x_i - \bar{x})(y_i - \bar{y})}{ \sqrt{\sum_{i=1}^{N} (x_i - \bar{x})^2} \sqrt{\sum_{i=1}^{N} (y_i - \bar{y})^2}},
\end{align}
where $\bar{x} = \frac{1}{N} \sum_{i=1}^{N}x_i$ and $\bar{y} = \frac{1}{N} \sum_{i=1}^{N}y_i$ correspond to the sample mean. The values of $r$ are within the interval $[-1,1]$, where $r=1(-1)$ corresponds to $100\%$ (anti-)linear correlation. $r=0$ corresponds to no linear correlation. Figure \ref{fig:pearson0} shows an example of two variables with no linear correlation and figure \ref{fig:pearson1} shows an example of two variables with linear correlation.

In general, it is not possible to conclude from the absence of linear correlation that two variables are independent. For example in case of two variables that follow a circular distribution, thus $x = r \cdot \cos \phi$ and $y = r \cdot \sin \phi$, the linear correlation coefficient is zero (see figure \ref{fig:pearson2}).

In HEP practice one should keep this limitation in mind as e.\,g. angular distributions can show a very small correlation coefficient to other variables but are not necessarily independent.

\section{Projections in subranges}
\label{sec:projections}
\begin{figure*}[t]
  \centering
  \subfigure[]{
    \includegraphics[width=0.3\linewidth, angle=0]{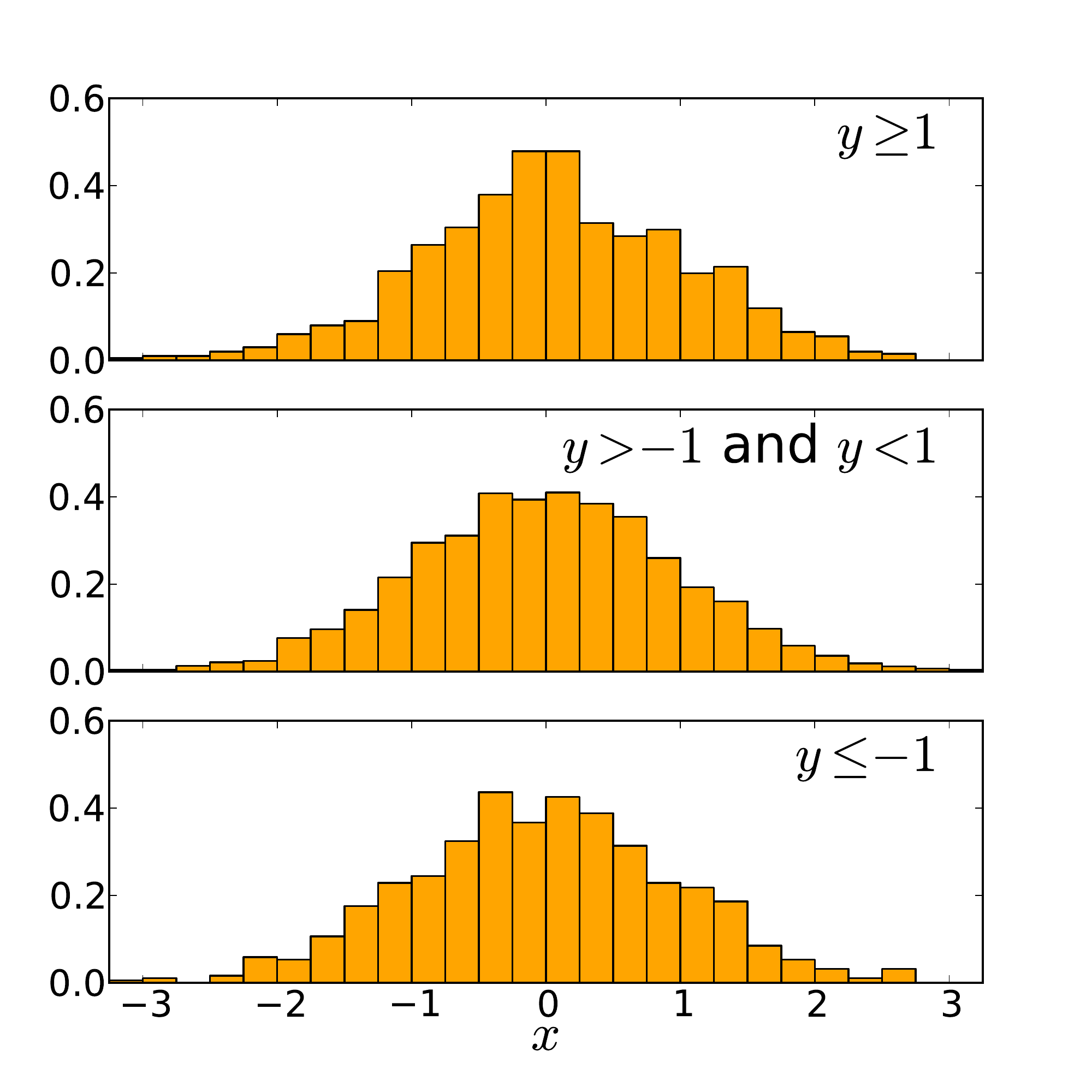}
    \label{fig:subrange0}
  }
  \subfigure[]{
    \includegraphics[width=0.3\linewidth, angle=0]{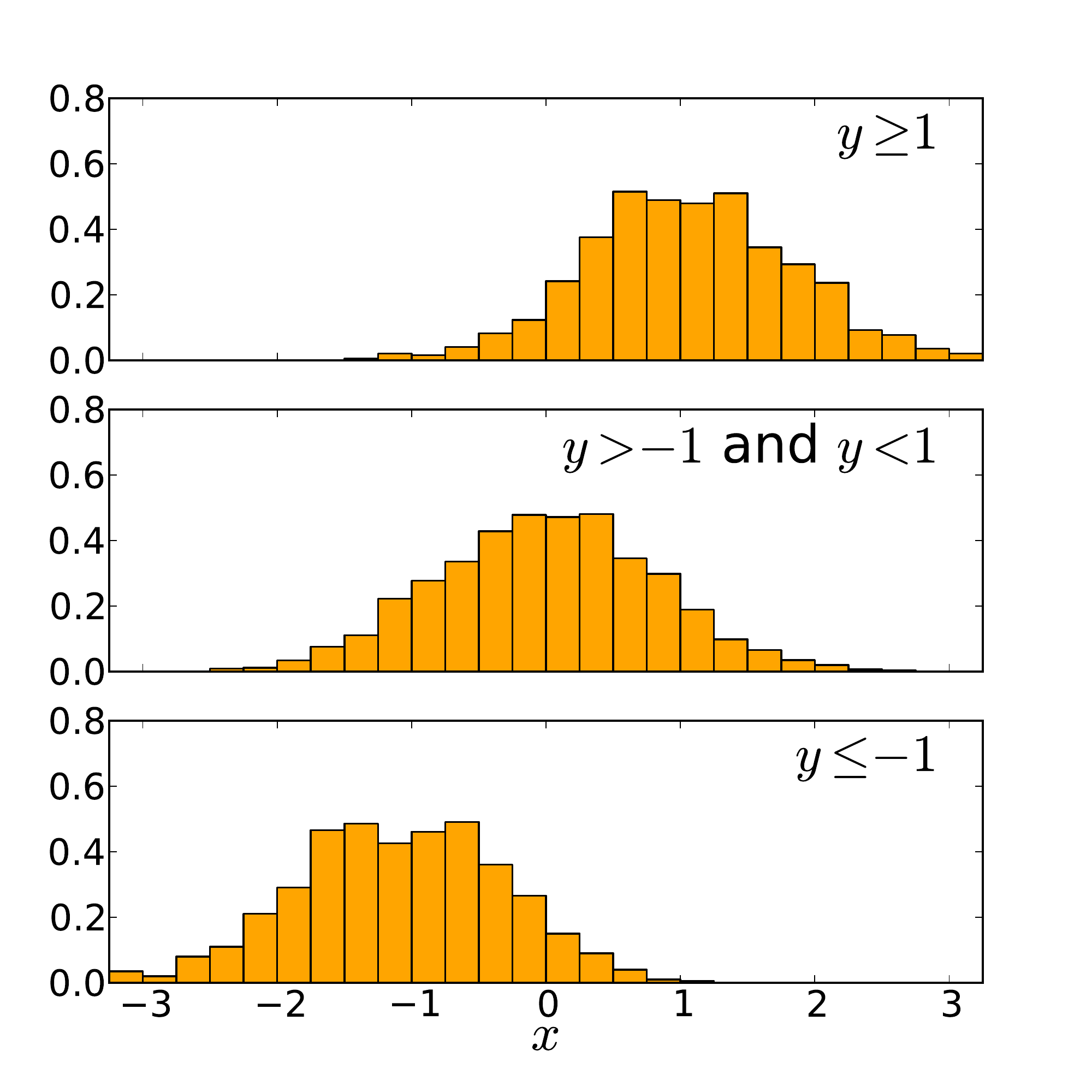}
    \label{fig:subrange1}
  }
  \subfigure[]{
    \includegraphics[width=0.3\linewidth, angle=0]{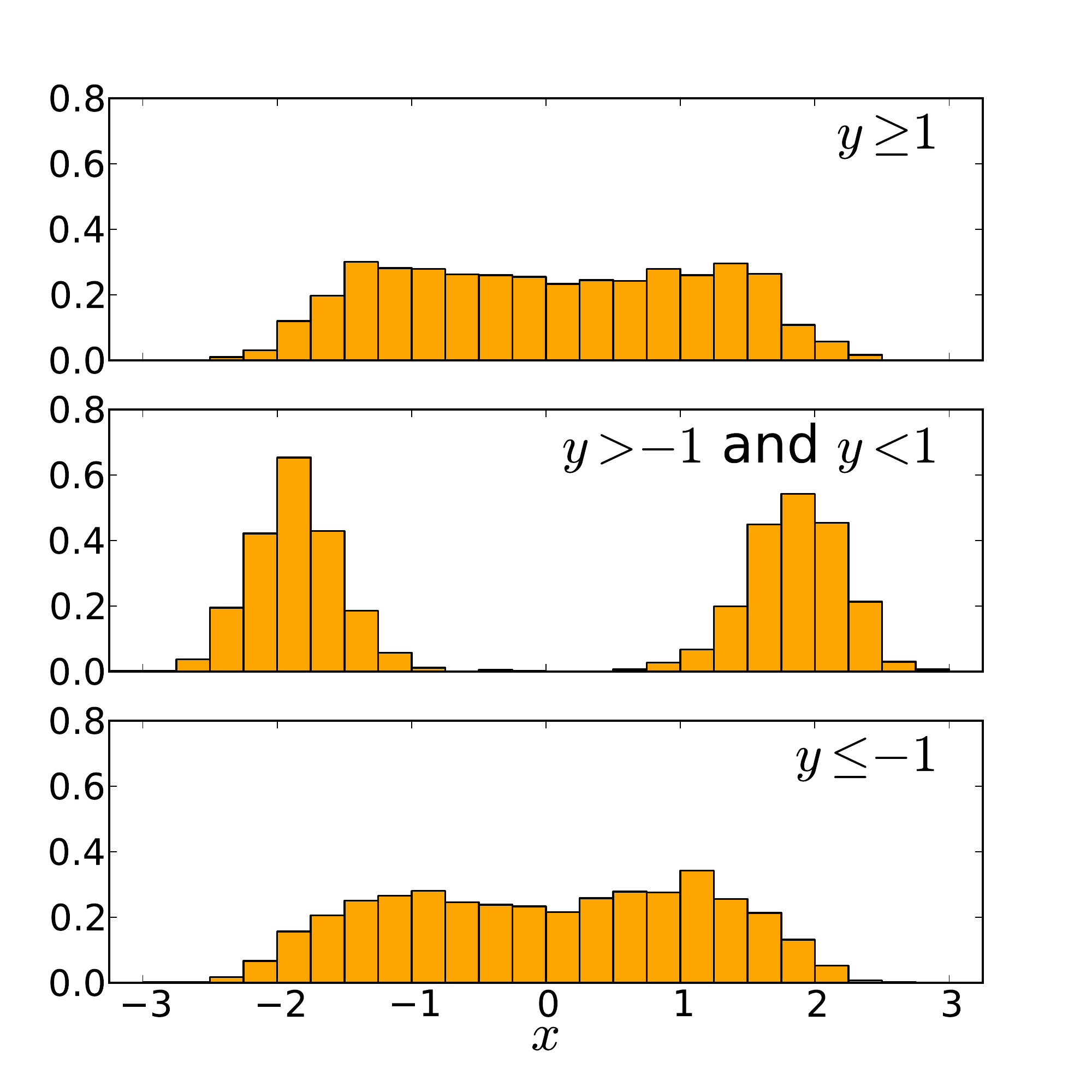}
    \label{fig:subrange2}
  }
  \caption{Normalized projections on variable $x$ in three different subranges of variable $y$ for the three data sets shown in figure \ref{fig:pearson}.}
  \label{fig:subrange}
\end{figure*}

To address the problem of dependencies between variables a common method in HEP is to look at projections of one variable in subranges of the other. In figure \ref{fig:subrange} three examples of this method are shown, using the same data sets that were introduced in figure \ref{fig:pearson}. In case of independent variables the three projections follow the same distribution. However, in general this method does not allow to conclude independence. One has to be aware of symmetry axes in the distribution. By choosing two bins with $y>0$ and $y<0$ instead of three, figure \ref{fig:subrange2} would lead to two similar distributions. By using an adequate number of bins this problem can be avoided in practical applications.

Another problem in practice is, that it might be hard to judge whether two variables are independent or not. Distributions might be very similar and compatible with each other within uncertainties or not. Statistical tests might be necessary to estimate their compatibility. In case of more than two variables it is also difficult to compare dependence and, e.\,g., sort them by their importance. The latter might be necessary to judge which dependencies should be described by a conditional PDF to improve the model. As these days multidimensional analyses with four, five or even more dimensions are becoming an important method, a reliable automatic procedure is desired.

\section{Hypothesis test for independence}
\label{sec:correlana}
Whereas the linear correlation coefficient is a quantitative measure of linear correlation, it can not be used to identify general dependence. On the other hand, projections in subranges can identify dependence but are difficult to compare or quantify without additional work.

\subsection{Copulas}
Copulas have been introduces in 1959 by Sklar to describe how a joint distribution function couples to its margins. Sklar's theorem states:

\textit{Let $S$ be a joint distribution function with margins $F$ and $G$. Then there exists a copula $C$ such that for all $x$,$y$ in $\mathbb{R}$,
\begin{equation}
S(x,y) = C(F(x),G(y)).
\label{eqn:sklar_theorem}
\end{equation}
If $F$ and $G$ are continuous, then $C$ is unique; otherwise, $C$ is uniquely determined on $\text{Ran} F \times \text{Ran} G$. Conversely, if $C$ is a copula and $F$ and $G$ are distribution functions, then the function $S$ defined by equation (\ref{eqn:sklar_theorem}) is a joint distribution function with margins $F$ and $G$.}

Sklar's theorem and more details on copulas can be found in \cite{Nelson}. A special copula is the unit copula $C(u,v) = u \times v$, which connects the marginal distributions of independent variables, as can be seen from equation (\ref{eqn:independence}).

\subsection{Hypothesis test for independence}
\label{sec:correlana_test}
We therefore present an algorithm that performs a test of the hypothesis whether in a given data set with $N$ events, two variables $x$ and $y$ are independent.

\begin{enumerate}
\item Determine the probability integral transforms $u=F(x)$ and $v=G(y)$ of variables $x$ and $y$. First sort the data in $x$ and $y$. The values of $u = I/N$ $(v = J/N)$, where $I(J)$ is the index of $x(y)$ in the sorted range, respectively, are then within the interval $[0,1]$. This is sometimes referred to as flattening the distribution.
\item Create a $n \times n$ histogram $H(u,v)$ with bins of equal size and fill it with all events. The number of bins $n$ should be chosen such that $N/n^2$ is large enough ($\gtrsim 25$). $H(u,v)$ corresponds to the empirical copula density.
\item In each bin of $H(u,v)$, if $x$ and $y$ are independent, we expect $e = N/n^2$ entries and the statistical uncertainty can be approximated by $\sigma_e = \sqrt{N/n^2}$ if the binning was chosen as suggested in step~1.
\item Compute the $\chi^2 = \sum_{i=1}^n \sum_{j=1}^n \frac{(h_{i,j} - e)^2}{\sigma_e^2}$, where $h_{i,j}$ is the content of the $(i,j)$\textsuperscript{th} bin of $H(u,v)$.
\item The probability of the data being consistent with a flat hypothesis and thus $x$ and $y$ being independent variables follows a $\chi^2$ distribution with $n^2 - (2n-1)$ degrees of freedom. By construction the number of degrees of freedom is reduced by $(2n-1)$ due to the flatness of the two marginal distributions.
\end{enumerate}

\begin{figure*}[t]
  \centering
  \subfigure[Probability $p = 0.1995$ corresponds to approximately $1.3\sigma$ significance.]{
    \includegraphics[width=0.3\linewidth, angle=0]{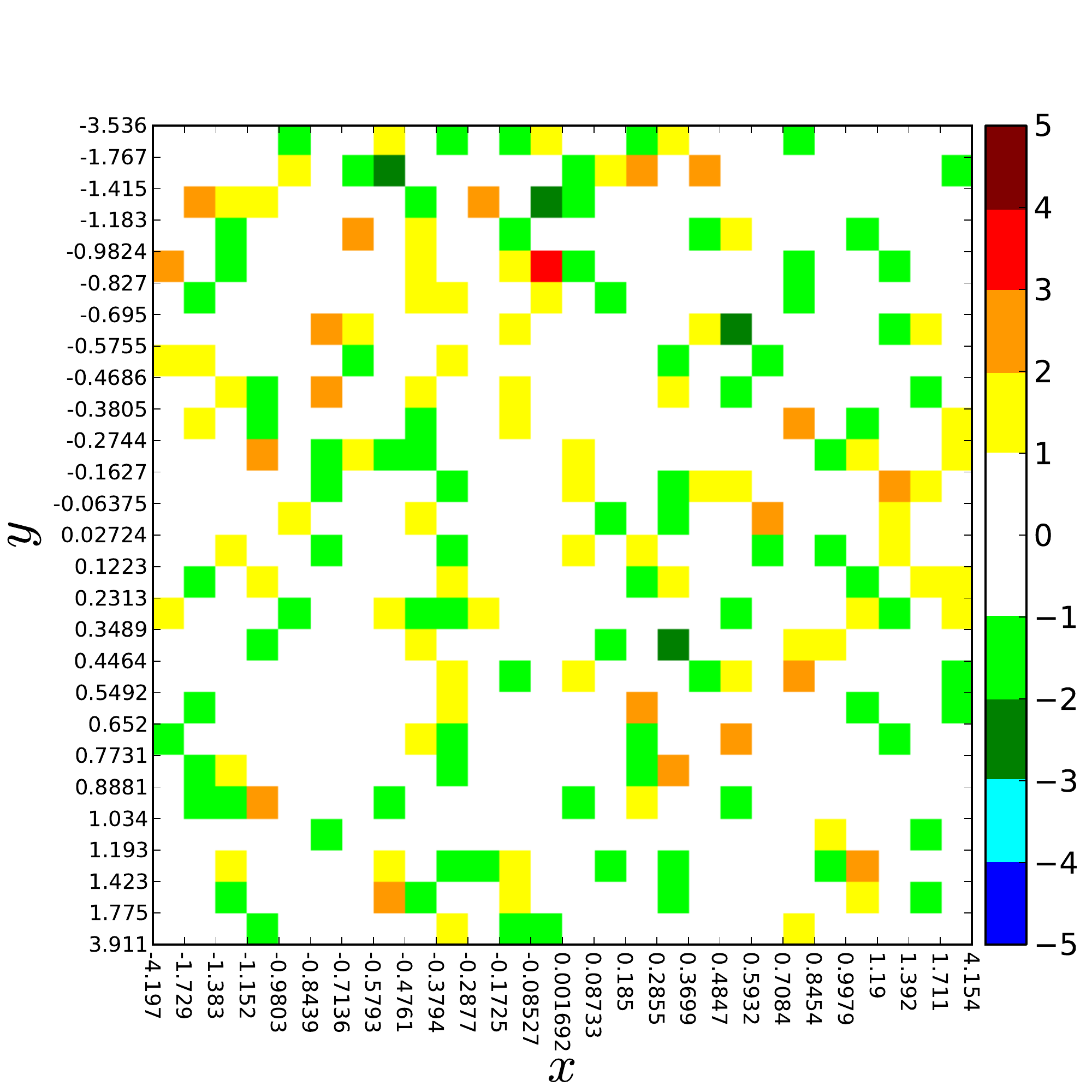}
    \label{fig:flat0}
  }
  \subfigure[Probability $p < 10^{-15}$ corresponds to more than $8\sigma$ significance.]{
    \includegraphics[width=0.3\linewidth, angle=0]{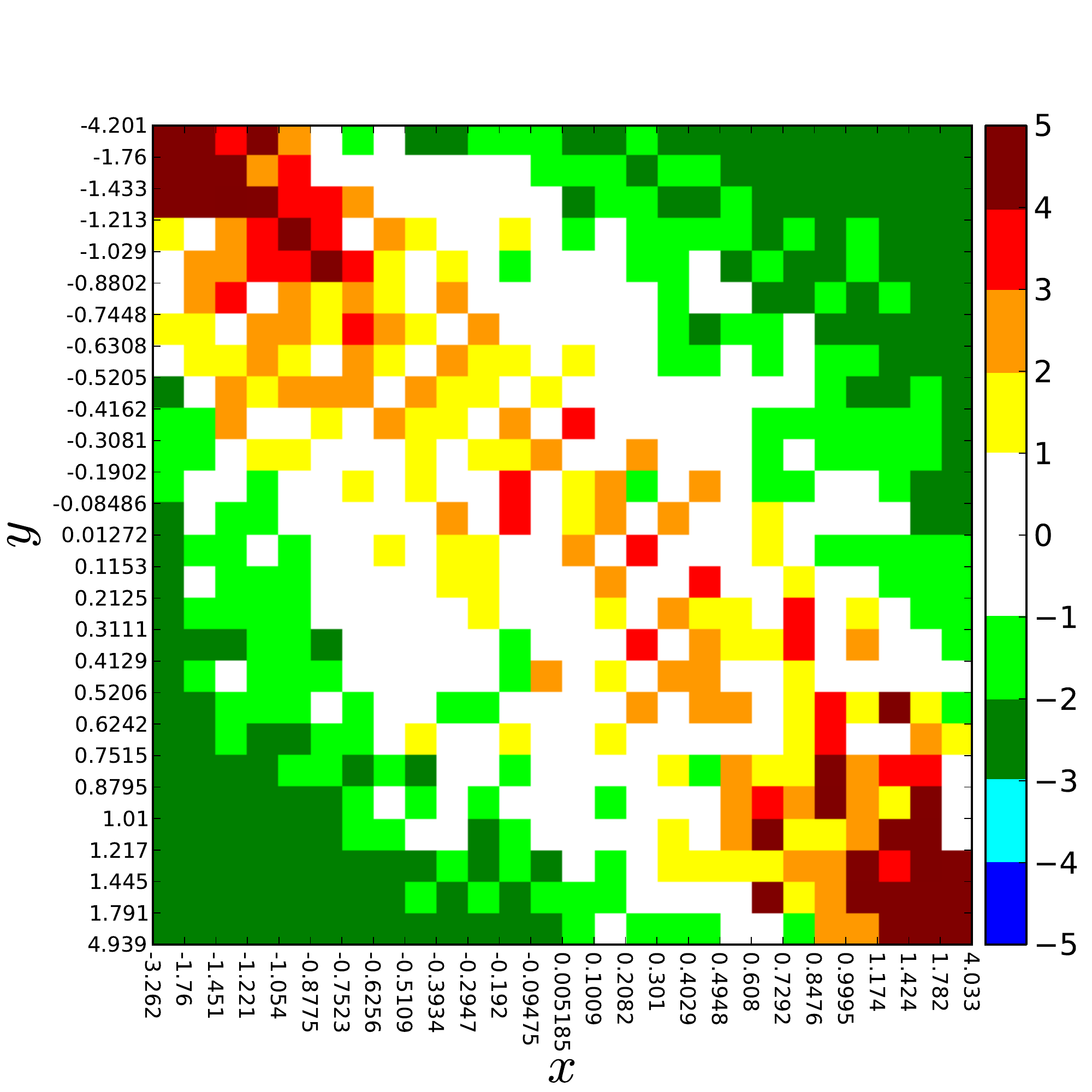}
    \label{fig:flat1}
  }
  \subfigure[Probability $p < 10^{-15}$ corresponds to more than $8\sigma$ significance.]{
    \includegraphics[width=0.3\linewidth, angle=0]{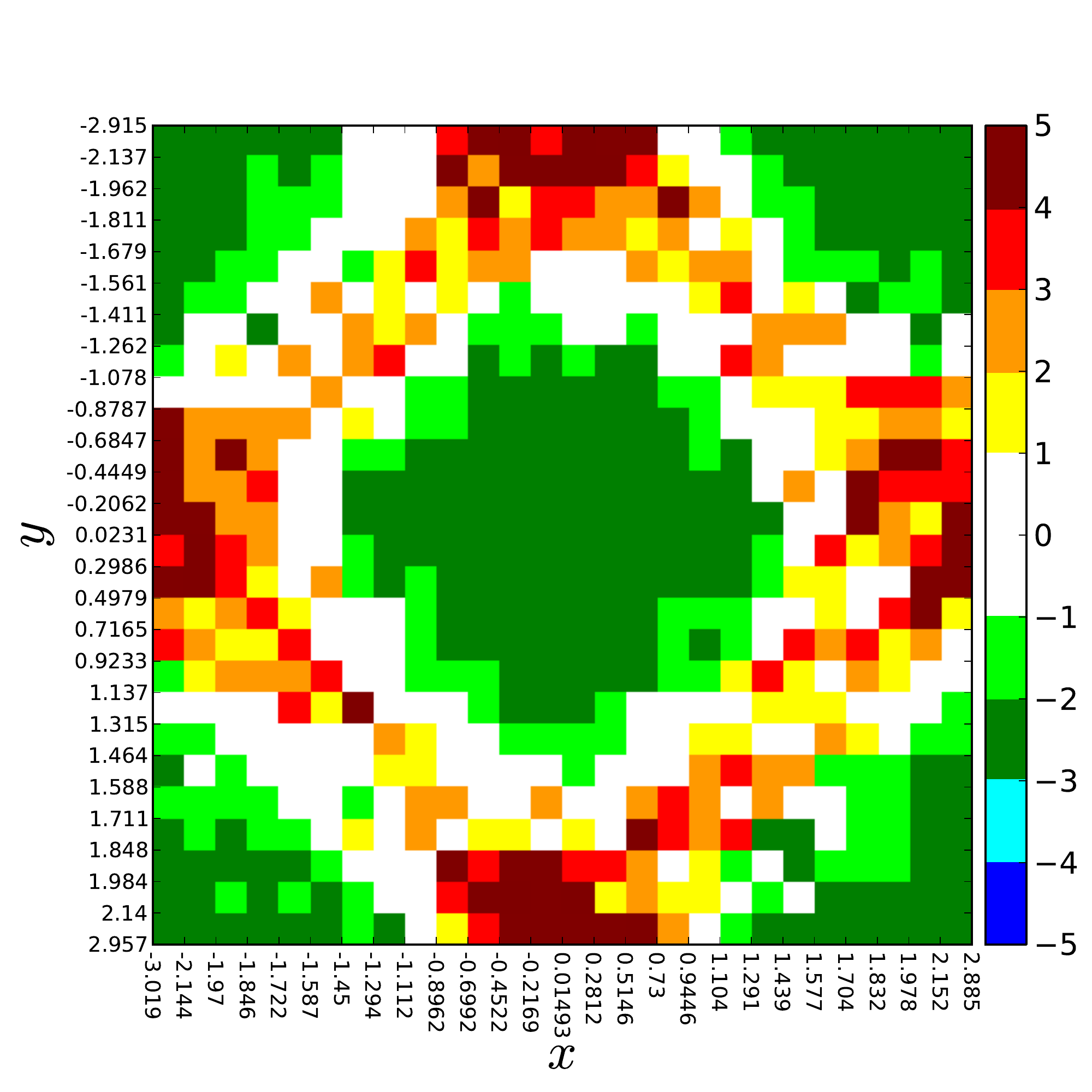}
    \label{fig:flat2}
  }
  \caption{Deviation in units of $\sigma_e$ for the histogram $H(u,v)$ from a flat distribution for the three data sets shown in figure \ref{fig:pearson}. The axis labels correspond to the untransformed (original) values of $x$ and $y$, which allow for a simpler interpretation than the values in $u$ and $v$. Resulting probabilities for the distribution being consistent with a flat distribution and transformation in units of standard deviations given below.}
  \label{fig:flat}
\end{figure*}

In short, the algorithm performs a test of $H(u,v)$ being consistent with the constant density $c(u,v) = \frac{\partial^2 C(u,v)}{\partial u \partial v}$ expected from the unit copula. The algorithm is able to identify any linear or non-linear dependence. The probability of the hypothesis can easily be compared among different pairs of variables in a multivariate data set with more than two variables. It also can be translated into the unit of standard deviations significance for the hypothesis that $x$ and $y$ are independent. See the section about significance tests in \cite[chap. 36.2.2]{PDG}. Examples of the resulting deviations from a flat distribution for histogram $H(u,v)$ are shown in figure \ref{fig:flat} for the data sets introduced in figure \ref{fig:pearson}.

The algorithm is very robust and delivers reliable results no matter whether variable values are located on a small interval or reach over several orders of magnitude as it is based on rank statistics.

Another feature of this algorithm is the fact that its output scales with the size of the data set. A dependence might be negligible for low statistics but significant for higher statistics. Imagine for example a chessboard like distribution. Neither the algorithm nor the maximum likelihood fit will be sensitive to this dependence with low statistics and a simple product of marginal distributions will describe the data. With increasing statistics this dependence will become more and more significant as the size of the bins decreases. Also the fit model will have to be adjusted once the dependence reaches a certain level.

\subsection{Practical application in HEP}
\label{sec:correlana_guide}
In practical HEP application of a multidimensional maximum likelihood analysis the output of the algorithm offers the experimentalist a reliable quantity for supporting the decision to choose a simple product approach in the construction of the PDF.

To verify that the approach is reasonable, a simulated data set with the same statistics as the real data set can be checked for any significant ($>5\sigma$) or evident ($>3\sigma$), if conservative, dependence. If available, e.\,g. for signal events, a larger simulated data set with 10 times the statistics could be checked to not have any significant dependencies. What can be done in case of dependencies will be briefly discussed in section \ref{sec:howto}.

It is however not recommended to check simulated data sets with e.\,g. 100 or 1000 times the statistics of real data, as it is sometimes available for signal events. Dependencies, which become significant only with these statistics, are negligible for a maximum likelihood analysis on real data statistics. Furthermore at such high statistics it might be questionable if the simulation has the proper level of accuracy to describe dependencies to that detail.

\section{CAT - A correlation analysis tool}
\label{sec:CAT}
A careful study of dependencies requires a non negligible amount of work. As we have shown, simple and fast methods such as the linear correlation coefficient, do not deliver a reliable result. We therefore developed a fully automatic tool, CAT, that performs an analysis for a given multivariate data set. Including such tool into the work-flow of a multidimensional maximum likelihood analysis could significantly shorten the amount time, which is necessary to understand the data sample. Currently the following methods, which partially have been discussed in this paper, are included:

\begin{enumerate}
\item Linear correlation coefficient
\item Profile plot of variable $x$ vs. variable $y$ and vice versa
\item Projections of variable $x$ in subranges of variable $y$ and vice versa
\item Hypothesis test of variable $x$ and $y$ being independent
\end{enumerate}

For a given data set with $n$ variables all methods are computed for all pairs of variables automatically. An analysis report file is created, which provides a nice visual review and numeric results.

CAT can be downloaded from \cite{CAT}. As input a comma separated value (CSV) file is used as such file can be produced easily from any type of user data format. A script to transform data from a flat ROOT \cite{ROOT} tuple to CSV is provided as this is expected to be the most common case for application in HEP. Beside this a script to generate some example random data sets with different dependencies is provided. CAT is licensed under the GPLv3 \cite{GPL}.

\section{How to deal with dependencies?}
\label{sec:howto}
Unfortunately, sometimes a product PDF is not a valid approach. Assuming three variables $x$,$y$ and $z$ and a significant dependence between $x$ and $y$, there are different possibilities. One simple possibility is of course to remove either $x$ or $y$ from the maximum likelihood analysis and perform e.\,g. a simple cut on it. A more complicated approach would be to perform the maximum likelihood analysis in bins of either $x$ or $y$. The latter can also be a first step to understand the dependence better and to finally describe the probability density function as conditional PDF and thus the model becoming $\mathcal{P}(x,y,z) = \mathcal{P}(x\vert y)\times \mathcal{P}(y) \times \mathcal{P}(z)$. Whichever method is chosen, dealing with dependencies can be a more complicated problem than identifying them. Even more important it is to be able to show that neglecting dependencies is a valid approach.

\section{Possible applications beyond maximum likelihood fits}
In this paper we compare the empiric copula density against the expected density from the unit copula to search for dependence. In principle comparisons of the empiric copula density can also be made against other copulas, which e.\,g. describe the expected density from standard model physics. Such an approach does not require any model assumptions about possible physics beyond the standard model. Similar approaches have e.\,g. been made with the Sleuth algorithm in \cite{Sleuth}.

Another possible application is the identification of good input variables for multivariate methods. A high dependence between the target variable and input variables is usually desired. Depending on the problem it might also be that variables that have a large dependence on a certain variable shall be excluded such that the multivariate method can not influence this specific variable. A multivariate method should for example not produce a peak in the mass distribution of background events, which can be avoided by removing variables that have a strong dependence on the mass. A widely used multivariate data analysis package is TMVA \cite{TMVA}, which is included in ROOT. Beside this, the NeuroBayes \cite{NeuroBayes} package, which was developed in HEP, has also found wide application among different experiments. A general review of multivariate methods and applications in HEP can be found in \cite{MLmethods}.

\section{Conclusion}
\label{sec:conclusion}
We have presented an algorithm that is able to quantify dependencies in multivariate data sets. The algorithm is able to deliver a reliable measure of dependence for supporting the product approach in multidimensional likelihood analyses. We have shown how to interpret its result in practice and we expect it to be a very useful method as these days more and more complicated and multidimensional analyses are carried out in HEP.

In addition a fully automatic tool, CAT, was presented that performs a comprehensive analysis for a given multivariate data set and creates an analysis report.

\end{document}